\documentclass[prb, showkeys,preprintnumbers]{revtex4}
\newcommand{\onvire}[1]{}

\usepackage{color}
\usepackage{amssymb}
\usepackage{graphicx}

\begin{document}

    \preprint{LAPTH-XXX/03}
    \title{Diffusion in a slab: different approaches}
    
    \author{R. Taillet} \email{taillet@lapp.in2p3.fr}
    \author{P. Salati}\email{salati@lapp.in2p3.fr}
    \affiliation{Laboratoire de Physique Th\'eorique LAPTH, Annecy--le--Vieux, 74941, France}
    \affiliation{Universit\'e de Savoie, Chamb\'ery, 73011, France}
    \author{D. Maurin} \email{dmaurin@discovery.saclay.cea.fr}
    \affiliation{SAp, CEA, Orme des Merisiers, 91191 Gif-sur-Yvette, France} 
    \author{E. Pilon}\email{pilon@lapp.in2p3.fr}
    \affiliation{Laboratoire de Physique Th\'eorique LAPTH, Annecy--le--Vieux, 74941, France}

\date{\today}

\begin{abstract}
Different approaches are presented to investigate diffusion from a point source in a slab delimited by two 
absorbing boundaries consisting of parallel infinite planes. These approaches enable to consider the effect 
of absorption at the boundaries as well as the possibility that the particles that diffuse react with the 
diffusive medium.
\end{abstract}

\maketitle


\section{Introduction}

The problem of steady-state diffusion from a source $q(\vec{r})$ is described by a quite simple equation
\begin{equation}
    K \Delta N(\vec{r}) = -q(\vec{r})
\end{equation}
where $K$ is the diffusion coefficient (homogeneous to a surface per unit time).
For a point source $q(\vec{r}) \equiv \delta^3(\vec{r})$ embedded in infinite space, the solution is also
simple, given by
\begin{equation}
    N(\vec{r}) = \frac{1}{4\pi K |\vec{r}|}
    \label{free_sol}
\end{equation}
For more complicated geometries of the diffusive volume, the solutions lose this simplicity. In some cases, 
it is helpful to use approaches different from brute-force resolution. We want to illustrate this point for 
the problem of diffusion in a slab delimited in the $z$ direction by two absorbing boundaries, located at $z=\pm 
L$ and imposing that $N(z=\pm L) = 0$.
We present four approaches, giving identical results when the conditions of validity overlap, taking also into 
account the possibility that the particles that diffuse can also be destroyed by reacting with the diffusive 
medium.
The equation then reads
\begin{equation}
    K\Delta N - \Gamma(\vec{r}) N = -q(\vec{r})
    \label{diffusion_stat}
\end{equation}
where the destruction rate $\Gamma(\vec{r})$ is related to the density $n(\vec{r})$ of the reacting medium, the 
reaction cross-section $\sigma$ and the velocity $v$ of the diffusing particle by
$\Gamma(\vec{r}) = n(\vec{r}) \sigma v$.
The authors first encountered this situation when studying the diffusion of cosmic rays emitted from sources 
in the galactic plane \cite{taillet03}. The galactic magnetic field has a stochastic component which is responsible for their 
diffusion, but also for their confinement in a volume which corresponds approximately to the geometry 
described above. The destruction term corresponds to the nuclear reactions (spallations) that may occur when 
these cosmic Rays cross the regions of the galactic disk where the nuclei of interstellar matter are present.
This is why we pay a particular attention to the situation where the sources,
the measurement and the destruction process are localized in the plane $z=0$.
\begin{figure}[!h]
    \centerline{\includegraphics[width=6cm]{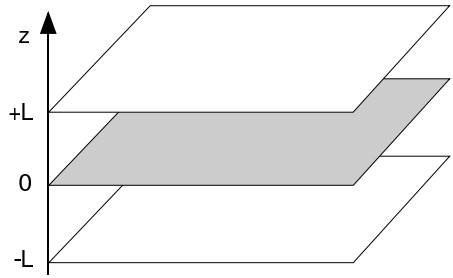}}
    \caption{Geometry of the diffusive slab. The central plane contains the matter on which 
    the diffusing particles can react. The upper and lower planes are absorbing boundaries, imposing a null 
    density.}
    \label{fig:schema_slab}
\end{figure}


\section{Steady-state solution using Fourier-Bessel transforms}
\label{sec:bessel}

\subsection{Diffusion equation}

Given the geometry of the diffusion volume and the source, it is easier to use cylindrical coordinates.
In all the following, the destruction term will not depend on the radial coordinate $r$, and we have
\begin{equation}
    \frac{d^2N}{dr^2} + \frac{1}{r} \frac{dN}{dr}
    + \frac{d^2N}{dz^2}  - 
    \frac{\Gamma(z)}{K} N = -\frac{q(r,z)}{K}
    \label{diffusion_equation}
\end{equation}
We develop $N(r,z)$ et $q(r,z)$ over the Bessel functions $J_0(k r)$, i.e.
\begin{equation}
    N(r,z) = \int_0^\infty dk\, k \, J_0(k r)  \, \tilde{N}(k,z) 
    \label{dvpt_bessel}
\end{equation}
with\footnote{These expressions are very similar to those obtained with a circular boundary, as given by 
Jackson, except that summations over Bessel functions become integrals, and with the substitutions
$1/J_1^2(\zeta_i) \rightarrow k\pi R/2$, $\sum_i \rightarrow  \int 
d(Rk/\pi)$ and $\zeta_i/R \rightarrow k$.}
\begin{equation}
    \tilde{N}(k,z) =  \int_0^\infty dr \, r \, J_0(kr) \, N(r,z) 
\end{equation}
The radial dependence is now encoded in the relative weight of the Bessel functions.
High values of $k$ correspond to finer details in the radial distribution, much like for usual Fourier 
transforms.
Inserting (\ref{dvpt_bessel}) into (\ref{diffusion_equation}), and using the fundamental property of Bessel 
functions
\begin{equation}
    J_0''(x) + \frac{1}{x} J_0'(x) = -J_0(x)
\end{equation}
we obtain
\begin{equation}
    -\int_0^\infty  dk \, k^3 \, J_0(kr)  \, \tilde{N}(k,z) 
    + \int_0^\infty dk \, k \, J_0(kr)\, 
    \left( \frac{d^2\tilde{N}(k,z)}{dz^2} - 
    \frac{\Gamma(z)}{K} \tilde{N}(k,z) \right) = -\frac{1}{K} \int_0^\infty dk \, k \, J_0(kr) \, \tilde{q}(k,z) 
\end{equation}
Using the property of orthonormalization
\begin{equation}
    \int_0^\infty dr \, r \, J_0(k_1r) J_0(k_2r) = \delta(k_1^2 - k_2^2)
\end{equation}
we select the equation for each mode $k$
\begin{equation}
    \frac{d^2\tilde{N}(k,z)}{dz^2} - 
    \left(\frac{\Gamma(z)}{K} + k^2 \right) \tilde{N}(k,z)
    = -\frac{\tilde{q}(k,z)}{K} 
\end{equation}

\subsection{Solution for a destructive plane}

\begin{figure}[!h]
    \centerline{\includegraphics[width=9cm]{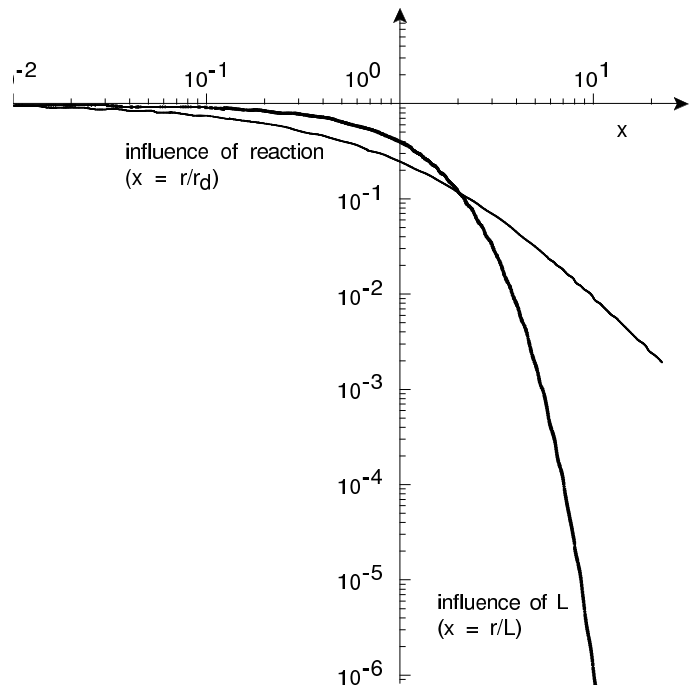}}
    \caption{Deviation from the free diffusion case $1/4\pi K r$, as a function of $r/L$ for the case without 
    reaction (thick curve: finite $L$, $\Gamma=0$) and of $r/r_d$ for the case without boundary 
    (thin curve: $L\rightarrow \infty$, finite $\Gamma$).}
    \label{fig:influence_L_rd}
\end{figure}

If destruction is confined to the plane $z=0$, we have $\Gamma(z)=\Gamma \delta(z)$.
We also consider a point-like source located in the disk, i.e. $q(r,z)=\delta(z) \delta(\pi r^2)$.
The equation to be solved is, for each mode $k$, 
\begin{equation}
    \frac{d^2\tilde{N}(k,z)}{dz^2} - 
    \left(\frac{\Gamma \delta(z)}{K} + k^2 \right) \tilde{N}(k,z)
    = -\frac{\delta(z) \tilde{q}(k)}{K} 
    \label{diffusion_spallation}
\end{equation}
In this expression, 
the combination $\Gamma/K \equiv 1/r_{\rm d}$ has the dimension of an inverse length, and the 
Bessel transform of a point-like source reads
\begin{equation}
    \tilde{q}(k) \equiv \int_0^\infty dr \, r \, J_0(kr) \, \delta(\pi r^2)
    = \frac{1}{2\pi}
\end{equation}
Outside of the disk ($z\neq 0$), the equation simplifies into
\begin{equation}
    \frac{d^2\tilde{N}(k,z)}{dz^2} -  k^2  \tilde{N}(k,z) = 0
\end{equation}
The solution has to be even in $z$ and satisfy the boundary conditions
$N(k,z=\pm L)=0$. Therefore, it is given by
\begin{equation}
    \tilde{N}(k,z\neq0) = \tilde{N}_0(k) \, \frac{\sinh \left\{ k (L-|z|) \right\} }{\sinh \left\{ k L \right\}}
    \label{sol1}
\end{equation}
The integration constant $\tilde{N}_0(k)$ is fixed by examining the solution in the plane $z=0$.
The derivatives of the solution (\ref{sol1}) for $z=0$ are defined only in terms of distributions. 
The computation may be made easier using the identity
\begin{equation}
    \exp |z| = \exp (-z) + 2 \Theta(z) \sinh z
\end{equation}
in the hyperbolic functions, as the derivative of the Heaviside distribution $\Theta(z)$ is the Dirac 
distribution $\delta(z)$.
This yields
\begin{equation}
    \tilde{N}(k,z\neq0) = \tilde{N}_0(k) \, \frac{\sinh \left\{ k (L-z) \right\} }{\sinh \left\{ k L \right\}}
    - 2 \tilde{N}_0(k) \theta(-z) \coth(kL) \sinh (kz)
\end{equation}
The second derivative reads
\begin{equation}
    \frac{d^2\tilde{N}(k,z)}{dz^2} = k^2 \, \tilde{N}(k,z)
    - 2 \tilde{N}_0(k) k \delta(z) \coth(kL)
\end{equation}
Inserting this last expression into (\ref{diffusion_spallation}) yields
\begin{equation}
    \tilde{N}_0(k) = \frac{1/2\pi}{\Gamma + 2 K k \coth(kL)}
\end{equation}
and  the final solution is
\begin{equation}
    N(r,z) = \int_0^\infty \frac{k\, J_0(kr) \, dk}{2\pi(\Gamma + 2 K k \coth(kL))} \, \frac{\sinh \left\{ k (L-|z|) \right\} }{\sinh \left\{ k L \right\}}
    \label{sol_J0}
\end{equation}
The diffusion process acts as a filtering in the Bessel space, the diffused density being related to the 
source by the transfer function
\begin{equation}
    T(k) = \frac{1}{\Gamma + 2 K k \coth(kL)}
\end{equation}
It is a low-pass filter. Diffusion tends to erase the small scale features.
The solution in the disk ($z=0$) can also be written under a form making apparent the 
correction to the free diffusion case (\ref{free_sol})
\begin{equation}
    N(r,z=0) = \frac{1}{4\pi Kr}
    \int_0^\infty \frac{kr\, J_0(kr) \, d(kr)}{r/2r_d + kr \coth(kL)} 
    \label{sol_J0_bis}
\end{equation}
This is illustrated in Fig.~\ref{fig:influence_L_rd}.

\subsection{Alternative formulations for a better convergence}

The integral involved in Eq.~(\ref{sol_J0_bis}) is of the form
\begin{equation}
        I[f] \equiv  \int_0^\infty J_0(x) \, f(x) dx
\end{equation}
When $f(\infty) =1$, as is the case up to a normalization in Eq.~(\ref{sol_J0_bis}) when $z=0$,
the slow decrease of the oscillations in the integrand makes the numerical computation of 
$I[f]$ quite tricky.
A few examples of manipulations which help to improve the numerical implementation of 
Eq.~(\ref{sol_J0_bis}) are given in the appendix.

\subsection{Solution for a short-lived species}

The same equation (\ref{diffusion_stat}) is also relevant to study the diffusion of unstable particles, with 
a lifetime $\tau$. The destruction rate is then homogeneous inside the diffusive volume, and given
by $\Gamma_u=1/\tau$.
It is straightforward to show, using the same procedure as above, that
in that case
\begin{equation}
    N(r,z=0) = \frac{1}{4\pi Kr}
    \int_0^\infty \frac{kr\, J_0(kr) \, d(kr)}{r/2r_d + S(k)r \coth(S(k)L)} 
    \label{sol_instable}
    \;\; \mbox{with} \;\;
    S(k) \equiv \sqrt{k^2 + \frac{\Gamma_u}{K}}
\end{equation}


\section{Time-dependent solution}
\label{sec:time}

A more precise description of the diffusion process may be obtained from the
time-dependent diffusion equation. 
This gives another formulation of the steady-state solution, as a series having better convergence 
properties. It also enables to take differently into account the case of decaying particles.

\subsection{The time-dependent diffusion equation}

The time-dependent diffusion equation reads
\begin{equation}
    \frac{\partial N}{\partial t} = 
    K \left\{ \frac{1}{r} \frac{\partial}{\partial r} \left(
    r \frac{\partial N}{\partial r} \right)+ \frac{\partial^2 N}{\partial z^2}
    \right\}
    - \Gamma \delta(z) N
    \label{eq_dif_prov}
\end{equation}
We seek the solution at $t>0$ such that $N(r,z,t=0) = \delta^3(\vec{r})$.
It is convenient to use the typical length 
$r_{\rm d}=K/\Gamma$ introduced above, or alternately its reciprocal $k_{\rm d}$,
so that
Eq.~\ref{eq_dif_prov} may be written as
\begin{equation}
    \frac{\partial N}{\partial (Kt)} = 
    \left\{ \frac{1}{r} \frac{\partial}{\partial r} \left(
    r \frac{\partial N}{\partial r} \right)+ \frac{\partial^2 N}{\partial z^2}
    \right\}
    - k_{\rm d}  \delta(z) N
\end{equation}
The diffusion process in the $z$ direction and in the radial 
direction are independent. 
As only pure diffusion occurs in the radial direction,
the density can be written as
\begin{equation}
    {\cal N}(r,z,t) = \frac{1}{4\pi K t} e^{-r^2/4Kt}
    \, N(z,t)
\end{equation}
where the function $N(z,t)$ satisfies
the time dependent diffusion equation along $z$
\begin{equation}
    \frac{\partial N}{\partial (Kt)} = 
    \frac{\partial^2 N}{\partial z^2}
    - k_{\rm d} \delta(z) N
\end{equation}

\subsection{Derivation}

First, we seek solutions of the form $N(z,t) = f(z) 
g(t)$, which separates the diffusion equation into
\begin{equation}
    g' = - \alpha g
    \;\;\; \mbox{and} \;\;\;
    - \alpha f = K f'' - \Gamma \delta(z) f
    \label{separation_t}
\end{equation}
where $\alpha$ must be positive in order to eliminate runaway 
solutions for $g$.
The equation on $f$ can be solved for $z>0$ and $z<0$ with the condition
$f(\pm L)=0$ as
\begin{equation}
    f(z) = A \sin \left\{k(L-|z|)\right\}
\end{equation}
where $k = \sqrt{\alpha/K}$. Derivation in the sense of the distribution, as in the previous section, yields
\begin{equation}
    K f''(z) = - \alpha f(z) - 2 K k A \delta(z) \cos kL 
\end{equation}
Inserting into (\ref{separation_t}) gives the condition
\begin{equation}
    2k \, \mbox{cotan} (kL) = - k_{\rm d}
    \label{eq_implicite_k}
\end{equation}
There is a infinite discrete set of $k_n$ satisfying the above 
condition, which gives the allowed values $\alpha_n=K k_n^2$ of $\alpha$.
For example, whit no reaction ($k_{\rm d}=0$), $k_n = (2n+1) \pi/2L$.
The general solution reads
\begin{equation}
    N(z,t) = \sum_{n=1}^\infty A_n e^{-\alpha_n t}
    \sin \left\{k_n (L-|z|)\right\}
\end{equation}
The functions $\sin \{k_n (L-|z|)\}$ form an orthogonal set, 
and it is found that
\begin{equation}
    \int_{-L}^L \sin \left\{k_n (L-|z|)\right\}
    \sin \left\{k_{n'} (L-|z|)\right\}\, dz
    = \delta_{n{n'}} c_n
\end{equation}
with
\begin{equation}
    c_n = L - \frac{\sin 2k_n L}{2k_n}
    \label{def_c_n}
\end{equation}
The $A_n$ are found by imposing that for $t=0$, the distribution is a 
dirac function, 
\begin{equation}
    \delta(z) = \sum_{n=1}^\infty A_n \sin \left\{k_n (L-|z|)\right\}
\end{equation}
Multiplying by $\sin \{k_m (L-|z|)\}$ and integrating over $z$ yields
\begin{equation}
    A_m =  c_m^{-1} \sin k_m L
\end{equation}
so that finally
\begin{equation}
    N(z,t) = \sum_{n=1}^\infty  c_n^{-1} e^{-\alpha_n t} \sin (k_n L)\, 
    \sin \left\{k_n (L-|z|)\right\}
    \label{sol_nonstat_1D}
\end{equation}
and
\begin{equation}
    {\cal N}(r,z,t) = \frac{1}{4\pi Kt} \exp \left(-\frac{r^2}{4Kt} \right)
    \sum_{n=1}^\infty  c_n^{-1} e^{-\alpha_n t} \sin (k_n L)\, 
    \sin \left\{k_n (L-|z|)\right\}
    \label{sol_nonstat_3D}
\end{equation}
The radial distribution in the disk is given by
\begin{equation}
    {\cal N}(r,z=0,t) = \frac{1}{4 \pi Kt} \exp \left(-\frac{r^2}{4Kt} \right)
    \sum_{n=1}^\infty c_n^{-1} e^{-k_n^2 Kt} \sin^2 (k_n L)
    \label{sol_nonstat_3D_disk}
\end{equation}
\begin{figure}[h!]
    \centerline{\includegraphics[width=5cm]{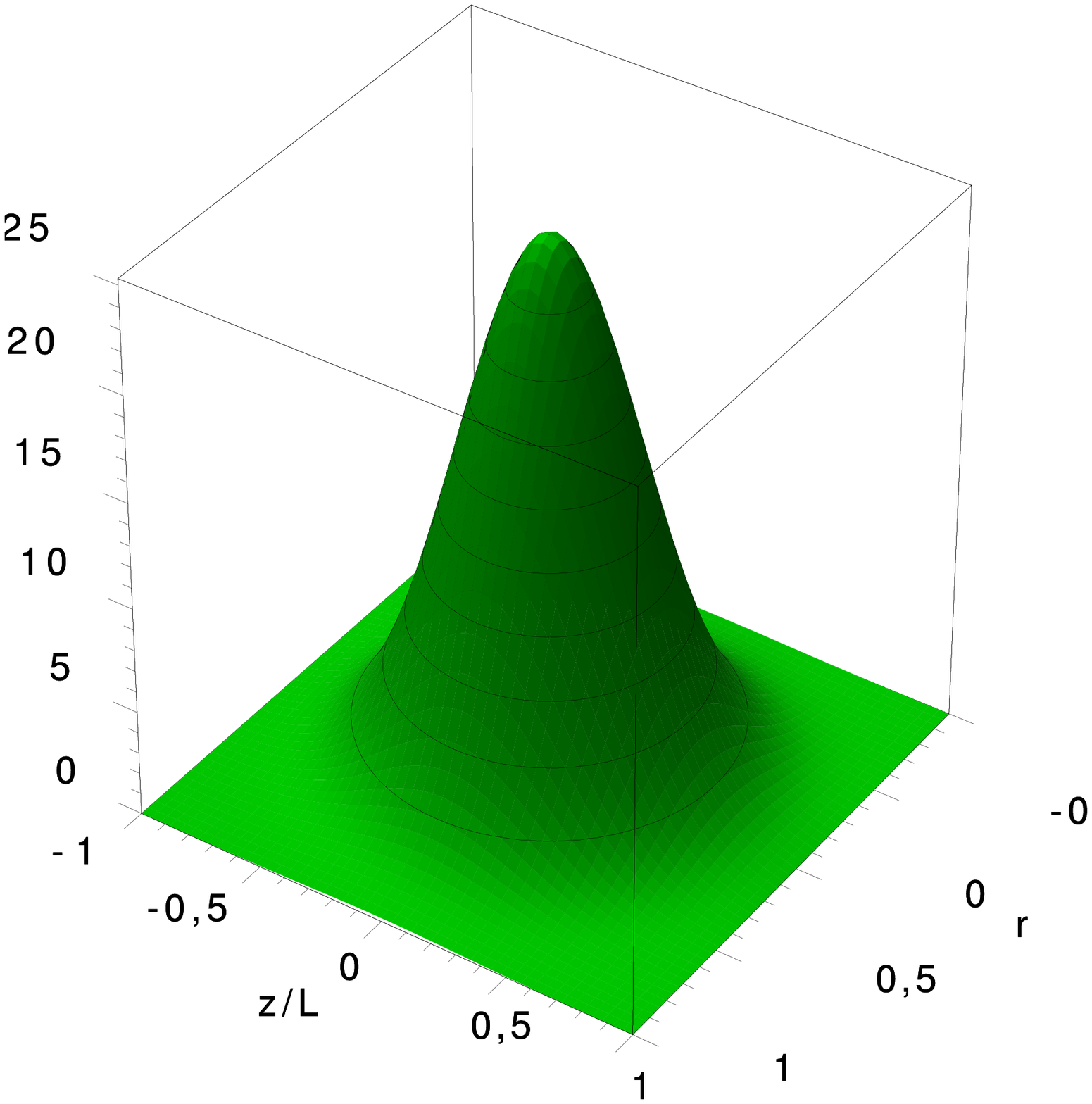}
    \includegraphics[width=5cm]{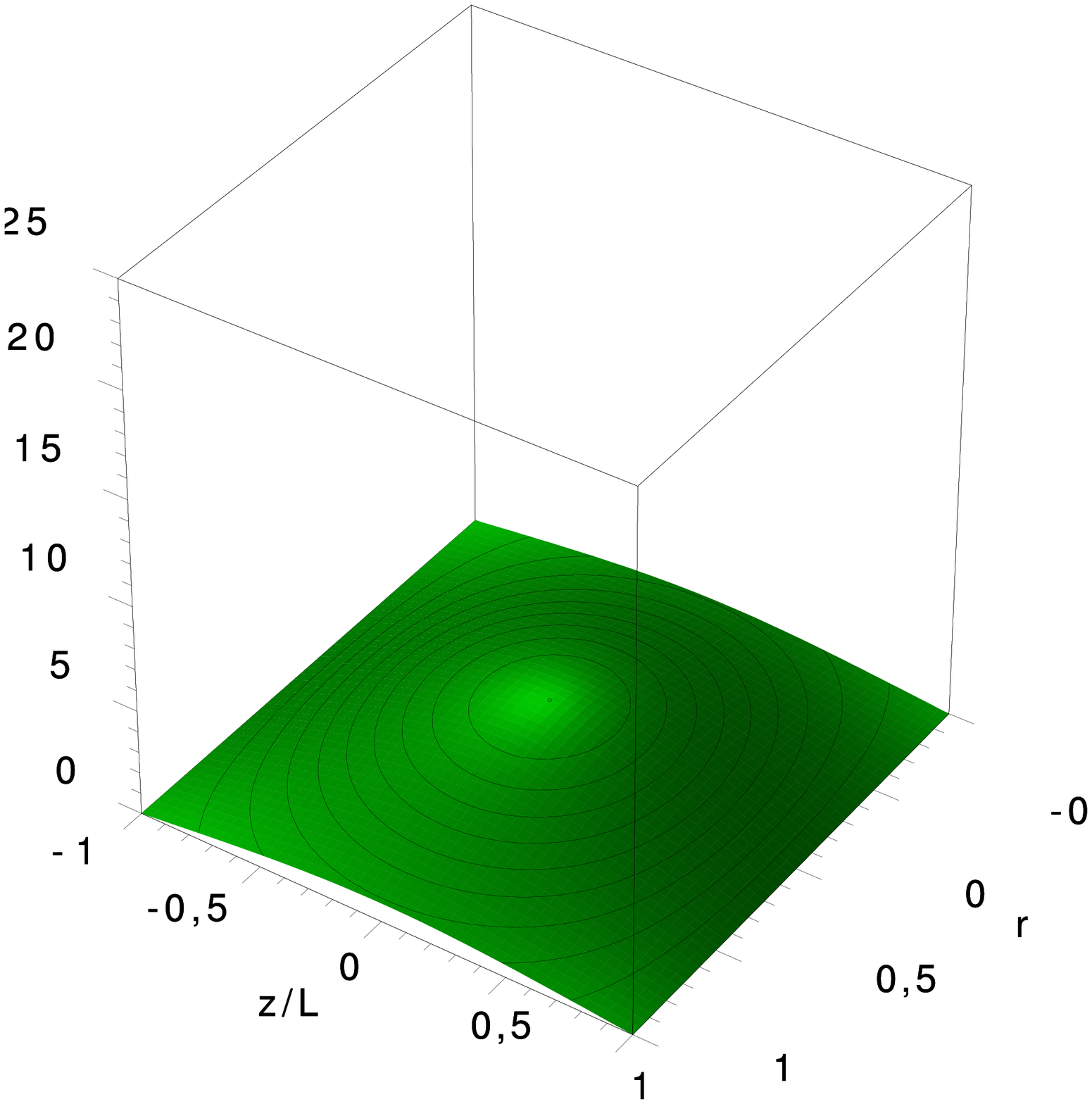}}
    \caption{Particle distribution as a function of $z/L$ and $r/L$, for $Kt=0.05 \, L$ and $Kt=0.2 \,L$.
    At early times, the distribution is close to the free case, as very few particles had time to reach the 
    boundary. At later times, the effect of absorption are more pronounced.}
\end{figure}

\subsection{Reformulation of the steady-state model}

\begin{figure}[!h]
    \centerline{\includegraphics[width=9cm]{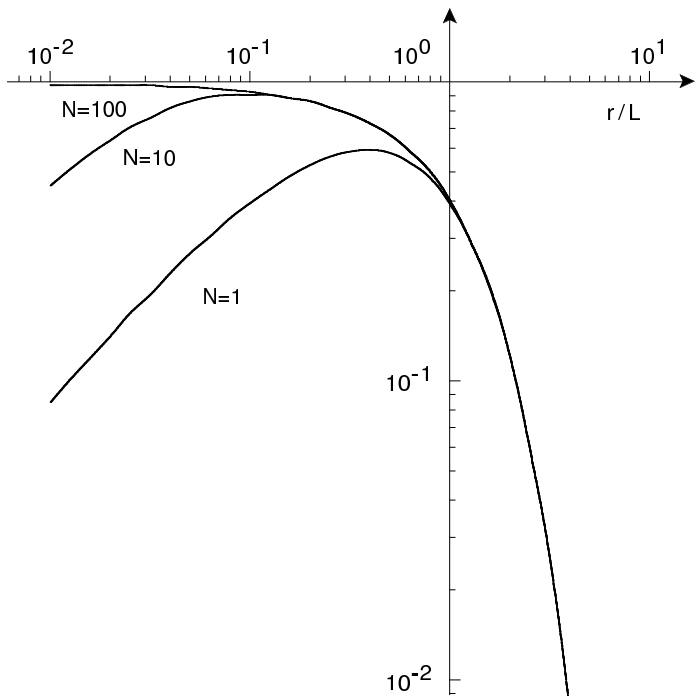}}
    \caption{Values of the correction to the free diffusion case as a function of $r/L$, for $\Gamma=0$, 
    computed with the series in $K_0$, truncated at different numbers of terms. The first term alone reproduces 
    quite well the profile for $r \gtrsim 2$.}
    \label{fig:somme_K0}
\end{figure}

The stationary regime results from the continuous superposition of 
solutions for instantaneous sources, so that the corresponding solution 
is given by
\begin{equation}
    {\cal N}_{\rm stat}(r,z) = \int_0^\infty {\cal N}(r,z,t) \, dt
\end{equation}
Using the identity (Gradshteyn \& Ryzhik 1980)
\begin{equation}
    \int_0^\infty \frac{dt}{t} \, e^{-\alpha t - \beta / t} = 2 K_0 \left(2 \sqrt{\alpha \beta} \right)
\end{equation}
involving the Bessel function of the third kind $K_0$, the integration of (\ref{sol_nonstat_3D}) yields
\begin{equation}
    {\cal N}_{\rm stat}(r,z) =   \frac{1}{2 \pi K} 
    \sum_{n=1}^\infty  c_n^{-1} K_0 \left( k_n r\right)
    \sin (k_n L)\, 
    \sin \left\{k_n (L-z)\right\}
\end{equation}
where the Bessel function of the third kind $K_0$ has been introduced.
The density in the disk is thus given by
\begin{equation}
    {\cal N}_{\rm stat}(r,z=0) =   \frac{1}{2 \pi K}
    \sum_{n=1}^\infty  c_n^{-1} K_0 \left( k_n r \right) \sin^2 (k_n L)
	\label{steady_0}
\end{equation}
This expression provides an alternative (but is exactly equivalent) to the usual Fourier Bessel 
expansion, using the $J_0$ functions. 
It is particularly well suited for sources well localized in space, 
like point-like sources, because the functions over which the 
development is performed (the $K_0$) do not oscillate. 
As a consequence, convergence of the series above is fast and the expression above provides a powerful 
alternative to compute the density for not too small values of $r/L$, as illustrated in Fig.~(\ref{fig:somme_K0})
for the case $\Gamma=0$, for which $\sin^2 k_n L=1$ and $c_n=L$, so that
\begin{equation}
    {\cal N}_{\rm stat}(r,z=0) =   \frac{1}{2 \pi KL}
    \sum_{n=1}^\infty   K_0 \left( \frac{2n+1}{2} \, \frac{\pi r}{L} \right) 
\end{equation}
This case is illustrated in Fig.~(\ref{fig:somme_K0}).

\subsection{The case of unstable particles}

When the particles can decay with a rate $\Gamma_u = 1/\tau$, the
previous time-dependent expressions must simply be corrected by a 
multiplicative factor $\exp(-\Gamma_u t)$, which amounts to
make the substitution $\alpha_n \rightarrow \alpha_n + \Gamma_u$
so that finally, the expression (\ref{steady_0}) is still valid
provided the substitution $k_n \rightarrow \sqrt{k_n^2 + \Gamma_u/K}$
is performed.

%


\section{Method of images}
\label{sec:images}

\subsection{Definition and result}

In the absence of destruction in the disk, a completely
different approach is provided by the elegant method of the images
inspired from electrostatics. In infinite space, the solution to
our diffusion problem is straightforward and is given by the
relation~(\ref{free_sol}) in the case of an initial Dirac
distribution at the source $S$. We furthermore would like to impose
the boundary condition according to which the cosmic-ray density $N$
vanishes on the plane $z = + L$ at any time. To do so, we can introduce
the virtual source $S'$ that is the image of the real source $S$
with respect to the boundary $z = + L$ acting like a anti-mirror.
The cosmic-ray densities which $S$ and its image $S'$ generate are
equal up to a relative minus sign that allows both contributions
to cancel out exactly on the boundary. 
In order to impose that the density also vanishes at $z = - L$, we can consider
the anti-image $S''$ of $S$ with respect to that lower boundary.
Because two anti-mirrors are now present at $z = + L$ and $z = - L$, an
infinite series of multiple images $\left\{ S_{n} \right\}$ arises.
They are aligned with the real source $S_{0} \equiv S$ in the vertical
direction and the position of $S_{n}$ is given by
\begin{equation}
     s_{n}  =  2  L n  + (-1)^{n} \, s
     \; .
\end{equation}
The virtual source $S_{n}$ results from $|n|$ reflections throughout
the mirrors and its production is affected by a sign $(-1)^{n}$ with
respect to the real source $S$. 
The distribution of sources within
the Galaxy is not perturbed by the presence of their virtual images
that are located outside the domain of interest. 
\begin{figure}[!h]
    \centerline{\includegraphics[width=9cm]{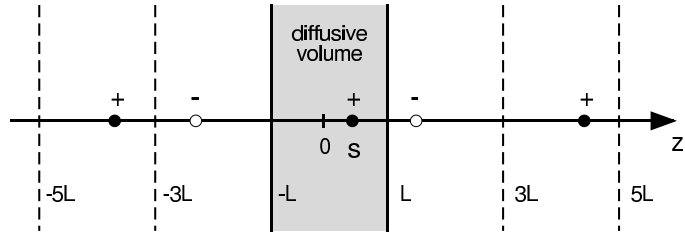}}
    \caption{Position of the virtual sources, images of the real sources through the anti-mirrors at $z=\pm L$.
    The black dots represent positive sources (they contribute like the real source) and white dots represent 
    negative sources.}
    \label{fig:images}
\end{figure}

We readily infer that in the presence
of boundaries, the time-dependent solution~(\ref{free_sol}) is modified into
\begin{equation}
     N_s (r,z,t) =
     \sum_{n = - \infty}^{+ \infty} \;
     \frac{(-1)^{n}}{(4 \pi K  t)^{3/2}} \;
     e^{- \left[ r^2 + \left( z - s_{n} \right)^{2}\right]/4 K t}
     \; .
     \label{images_solution_1}
\end{equation}
which can be rewritten, decoupling the diffusion processes along the vertical axis and in the radial 
direction, as
\begin{equation}
     N_s (r,z,t) =
     \frac{e^{- r^2/4 K t}}{4 \pi K t} \times
     \left\{ n \left\{ s,0 \to z,t \right\} =
     \sum_{n = - \infty}^{+ \infty} \;
     \frac{(-1)^{n}}{\sqrt{4 \pi K  t}} \;
     e^{- \left( z - s_{n} \right)^{2}/4 K t}
     \right\}
     \; .
     \label{images_solution}
\end{equation}
The steady-state solutions given by
\begin{equation}
     N_s (r,z,t) = 
     \sum_{n = - \infty}^{+ \infty} \;
     \frac{(-1)^{n}}{4 \pi K  \sqrt{r^2 + (z-s_n)^2}} 
     \label{images_solution_2}
\end{equation}
\begin{figure}[!h]
    \centerline{\includegraphics[width=9cm]{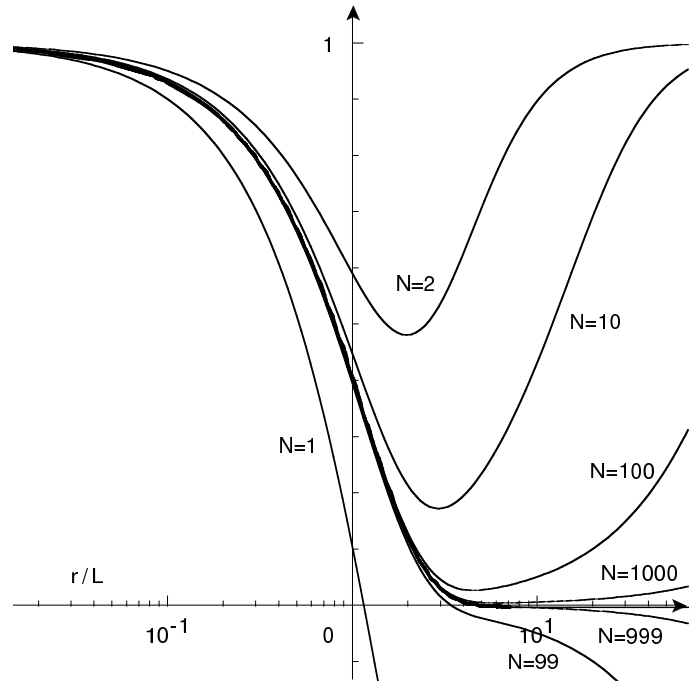}}
    \caption{Density in the plane $z=0$, as a function of $r/L$, and normalized to the free diffusion case. 
    The dots represent the exact solution, and the thin lines show the result given by the series of images, 
    for several truncatures of the sum.
    At very large distances ($r \gg L$) and in the case of a small number
    of sources, these are felt as a single source (positive or negative, 
    depending on the parity of $N$), so that the density tends to plus or minus the free diffusion case.}
    \label{fig:convergence_images}
\end{figure}

\subsection{Equivalence with the first approach}

The expression (\ref{sol_J0_bis}) can be easily transformed into (\ref{images_solution_2}), for $z=0$ 
in the case $\Gamma=0$.
Indeed, in the expression
\begin{equation}
    N(r,z) = \frac{1}{4\pi Kr} \, 
    \int_0^\infty J_0(kr) \, \tanh(kL) \, d(kr)
\end{equation}
using the development
\begin{equation}
    \tanh(kL)  = \left( 1-e^{-2kL} \right) \, \left( 1+e^{-2kL} \right)^{-1}
    = \left( 1-e^{-2kL} \right) \, \sum_{n=0}^\infty (-1)^n \, e^{-2nkL} 
\end{equation}
a few steps lead to
\begin{equation}
    N(r,z) = \frac{1}{4\pi Kr} \, \left\{
    1 +
    2\sum_{n=1}^\infty \int_0^\infty J_0(kr) \,  (-1)^n \, e^{-2nkL}  \, d(kr)
    \right\}
\end{equation}
Which finally gives the expression (\ref{images_solution_2}), 
using $\int_0^\infty J_0(ax)e \, \exp(-bx) dx = (a^2 + b^2)^{-1/2}$.

\subsection{Equivalence with the former approach}

The diffusion from the single source $S$ within a slab
-- on the boundaries of which the cosmic-ray density vanishes -- amounts
simply to the diffusion in infinite space from the series of sources
$S_{n}$. Along the vertical direction, the initial distribution is
\begin{equation}
     N_s(z,t) =
     {\displaystyle \sum_{n = - \infty}^{+ \infty}} \;
     (-1)^{n} \; \delta \left( z - s_{n} \right)
     \;.
     \label{sources_initial}
\end{equation}
Because that distribution is periodic, its Fourier transform which
we define by
\begin{equation}
     N(k) = \int_{- \infty}^{+ \infty} \,
     n \left\{ s,0 \to z,0 \right\} \;
     e^{- i k z} \,dz
     \; ,
\end{equation}
is composed of a discrete series of modes $k$. It may actually be expressed as
\begin{equation}
     N(k) =  \sum_{n = - \infty}^{+ \infty}\,
     (-1)^{n} \, e^{ - i k s_{n}} \equiv
     \left\{ e^{ - i k s}  -
     e^{2 k L} e^{i k s} \right\} \;
     \left\{ {\cal S} = \sum_{n = - \infty}^{+ \infty}\,
     e^{- 4 i n k L} \right\}
    \label{blabla}
     \; .
\end{equation}
After some straightforward algebra, very similar to what is done in optics to compute the
diffraction from an infinite diffracting grid, the sum ${\cal S}$ is transformed into
\begin{equation}
     {\cal S} =  \frac{\pi}{2L}\;
     \sum_{n = - \infty}^{+ \infty}\,
     \delta \left( k  -  n {\displaystyle \frac{\pi}{2L}} \right)
     \; .
\end{equation}
We conclude that the initial vertical distribution
$n \left\{ s,0 \to z,0 \right\}$ contains a series of modes with discrete
wavevectors $k = n  \pi / 2 L$. Inserting this expression into (\ref{blabla}), we see that
odd values of $n = 2p - 1$ are
associated to modes for which $k_{p} L = \left( p - 1/2 \right)  \pi$
and contribute a factor
\begin{equation}
     N \left( k_{p} \right) = \frac{\pi}{L} \, \cos k_{p} s
     \; ,
\end{equation}
whereas even values of $n = 2p$ lead to wavevectors $k'_{p}$ such that
$k'_{p} L = p  \pi$ and to
\begin{equation}
     N \left( k'_{p} \right) = -  i \frac{\pi}{L} \, \sin k'_{p} s
     \; .
\end{equation}
The initial cosmic-ray distribution in the vertical direction
$n \left\{ s,0 \to z,0 \right\}$ may be expressed as a Fourier series on
the various odd and even modes $p$ which we have just scrutinized
\begin{equation}
      N_s(z,t)=
     \sum_{p = 1}^{+ \infty} \;
     \left\{
     \frac{1}{L} \, \cos \left( k_{p} s \right) \, \cos \left( k_{p} z \right)
     +
     \frac{1}{L} \, \sin \left( k'_{p} s \right) \, \sin \left( k'_{p} z \right)
     \right\}
     \; .
\end{equation}
Because each Fourier mode $k$ exponentially decays in time like
$\exp \left( -  K k^{2} t \right)$ as a result of diffusion,
the initial distribution subsequently evolves into
\begin{equation}
     N_s(z,t) =
     \sum_{p = 1}^{+ \infty} \left\{
     \frac{e^{\displaystyle - \, K k_{p}^{2} t}}{L}
     \, \cos \left( k_{p} s \right) \, \cos \left( k_{p} z \right)
     +
     \frac{e^{\displaystyle - \, K {k'}_{p}^{2} t}}{L}
     \, \sin \left( k'_{p} s \right) \, \sin \left( k'_{p} z \right) \right\}
     \; .
\end{equation}
When the source is iat the origin ($s=0$), the density at the origin ($z=0$) is given by
\begin{equation}
     n \left\{ s,0 \to z,t \right\} =
     \sum_{p = 1}^{+ \infty}\;
     \frac{e^{\displaystyle - \, K k_{p}^{2} t}}{L}
     \, \cos \left( k_{p} z \right) \; .
\end{equation}
which, when radial diffusion is taken into account, is equivalent to the expression~(\ref{sol_nonstat_3D}) 
in the case of no reaction (then $\sin k_nL=1$ and $\cos k_nL=0$).


\section{Random walk approach}
\label{sec:random_walk}

It is well known that diffusion is closely related to random walks. The density at each point is related to the
number of stochastic path which reach this point, from the source.
When a boundary is present at $z=\pm L$, the paths that would go beyond this boundary do not contribute 
anymore to the density and must be discarded.
When destruction can occur in the disk, the paths that cross the disk should be attributed a lower weight.
In this section, we investigate separately these two effects.

\subsection{Probability of not escaping}

\begin{figure}[bt!]
\centerline{\includegraphics[width=9cm]{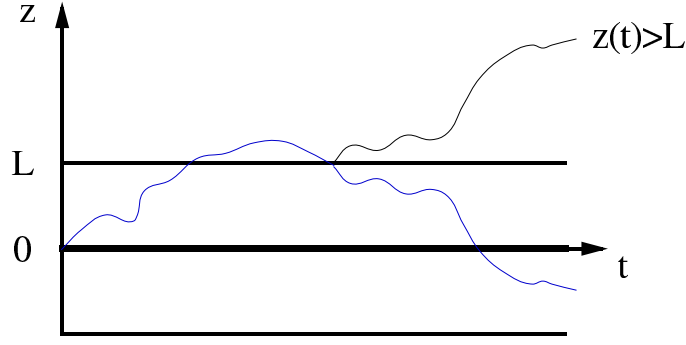}}
\caption{For each path such that $\mbox{max} \geq L$, there are two paths, one of which satisfies
$z(t) \geq L$ but not the other.}
\label{fig:marche_hasard_reflexion}
\end{figure}

\begin{figure}[bt!]
    \centerline{\includegraphics[width=9cm]{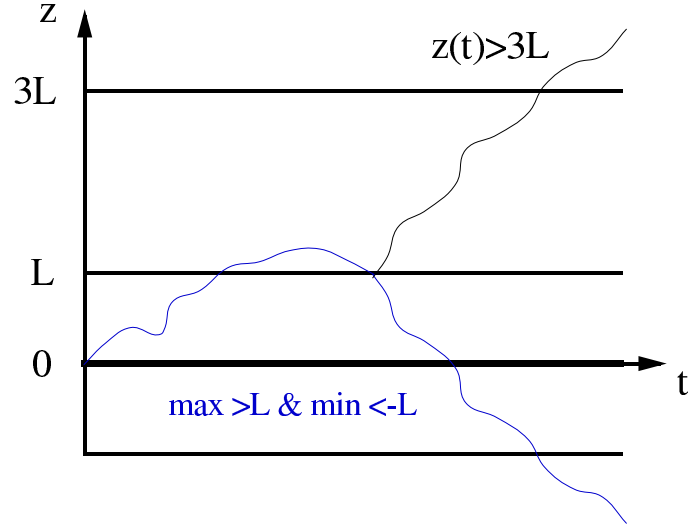}}
    \caption{For each path such that $\mbox{max} \geq L$ and
    $\mbox{min} \leq -L$, there is a symmetric path satisfying $z(t) \geq 3L$.}
    \label{fig:marche_hasard_reflexion2}
\end{figure}

We first compute the probability that a particle emitted at time $t=0$ in the disk
has not reached the boundary at a further time $t$.
At this time, position $z$ is given by a random walk of duration $t$.
The probability we seek is given by
\begin{equation}
    {\cal P} \left\{ (\mbox{max} < L) \cap (\mbox{min} > -L) \right\}
    = 1 - {\cal P} \left\{ (\mbox{max} \geq L) \cup (\mbox{min} \leq -L) \right\}
\end{equation}
where max and min are the maximum and minimum $z$ reached by the random walk.
From elementary statistics,
\begin{equation}
    {\cal P} \left\{ (\mbox{max} \geq L) \cup (\mbox{min} \leq -L) \right\}
    = {\cal P} \left\{ \mbox{max} \geq L \right\}+
    {\cal P} \left\{ \mbox{min} \leq -L \right\}-
    {\cal P} \left\{ (\mbox{max} \geq L) \cap (\mbox{min} \leq -L) \right\}
    \label{pt_depart}
\end{equation}
By symmetry, ${\cal P} \left\{ \mbox{max} \geq L \right\}=
{\cal P} \left\{ \mbox{min} \leq -L \right\}$. Furthermore, using the principle of reflection, it is 
straightforward to show that (see figure~\ref{fig:marche_hasard_reflexion}) 
\begin{equation}
    {\cal P} \left\{ \mbox{max} \geq L \right\} = 2 
    {\cal P} \left\{ z(t) \geq L \right\}
\end{equation}
It is also apparent that (see figure~\ref{fig:marche_hasard_reflexion2})
\begin{equation}
    {\cal P} \left\{ (\mbox{max} \geq L) \cap (\mbox{min} \leq -L) \right\}
    = {\cal P} \left\{ (\mbox{max} \geq 3L) \cup (\mbox{min} \leq -3L) \right\}
\end{equation}
so that finally
\begin{equation}
    {\cal P} \left\{ (\mbox{max} < L) \cap (\mbox{min} > -L) \right\}
    = 1 - 4{\cal P} \left\{ z(t) \geq L \right\}+
    {\cal P} \left\{ (\mbox{max} \geq 3L) \cup (\mbox{min} \leq -3L) \right\}
\end{equation}
Now, the same reasoning with Eq.~(\ref{pt_depart}) applied to the last term yields
\begin{equation}
    {\cal P} \left\{ (\mbox{max} < L) \cap (\mbox{min} > -L) \right\}
    = 1 - 4{\cal P} \left\{ z(t) \geq L \right\}
    + 4{\cal P} \left\{ z(t) \geq 3L \right\}
    - 4{\cal P} \left\{ z(t) \geq 5L \right\}
    + \ldots
\end{equation}

\subsection{Probability of not escaping for particles reaching the disk}

\begin{figure}[bt!]
\centerline{\includegraphics[width=9cm]{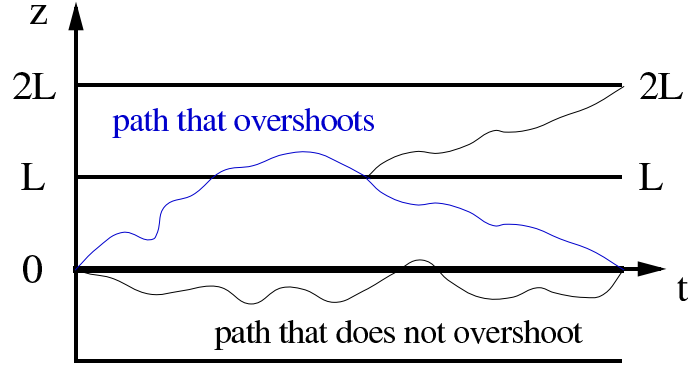}}
\caption{If a path such that $z(t)=0$ wanders beyond the upper boundary, then 
there is a symmetric path that satisfies $z(t)=2L$. 
The probability that the path goes beyond this boundary is thus given by
the ratio rapport ${\cal P} \left\{ z(t) = 2 L \right\}/{\cal P} \left\{ z(t) = 0 \right\}$.}
\label{fig:marche_hasard_L1}
\end{figure}

The problem we addressed in the previous sections was a bit different, though,
as we were interested in particles reaching a given point, for example in the disk. 
The same reasoning as before can be used, and the probability
that a particle reaching the disk at time $t$ has not wandered farther than the boundaries is given by
\begin{equation}
    {\cal P}_d \left\{ (\mbox{max} < L) \cap (\mbox{min} > -L) \right\}
    = 1 - {\cal P}_d \left\{ (\mbox{max} \geq L) \cup (\mbox{min} \leq -L)\right\}
\end{equation}
where we have used the compact notation for conditional probability
\begin{equation}
    {\cal P}_d \left\{ {\rm event} \right\} \equiv
    {\cal P} \left\{ {\rm event} |z(t)=0\right\}
\end{equation}
Exactly as before, the use of symmetry yields
\begin{equation}
    {\cal P}_d \left\{ (\mbox{max} \geq L) \cup (\mbox{min} \leq -L) \right\}
    = 2{\cal P}_d \left\{ \mbox{max} \geq L \right\}-
    {\cal P}_d \left\{ (\mbox{max} \geq L) \cap (\mbox{min} \leq -L) \right\}
    \label{pt_depart2}
\end{equation}
Now, the principle of reflection yields a different result for the conditional probability
(see figure~\ref{fig:marche_hasard_L1})
\begin{equation}
    {\cal P} \left\{ \mbox{max} \geq L |z(t)=0\right\} =  
    \frac{{\cal P} \left\{ z(t) = 2 L \right\}}
    {{\cal P} \left\{ z(t) = 0 \right\}}
\end{equation}
As before,
\begin{equation}
    {\cal P}_d \left\{ (\mbox{max} \geq L) \cap (\mbox{min} \leq -L) \right\}
    = {\cal P}_d \left\{ (\mbox{max} \geq 3L) \cup (\mbox{min} \leq -3L) \right\}
\end{equation}
so that finally
\begin{equation}
    {\cal P}_d \left\{ (\mbox{max} < L) \cap (\mbox{min} > -L) \right\}
    = 1 - 2\frac{{\cal P} \left\{ z(t) = 2 L \right\}}
    {{\cal P} \left\{ z(t) = 0 \right\}}+
    {\cal P}_d \left\{ (\mbox{max} \geq 3L) \cup (\mbox{min} \leq -3L) \right\}
\end{equation}
The same reasoning, applying equation~(\ref{pt_depart2}) to the last term,
gives
\begin{equation}
    {\cal P} \left\{ (\mbox{max} < L) \cap (\mbox{min} > -L) |z(t)=0\right\}
    = 1 - 2\frac{{\cal P} \left\{ z(t) = 2 L \right\}}
    {{\cal P} \left\{ z(t) = 0 \right\}}
    + 2\frac{{\cal P} \left\{ z(t) = 4 L \right\}}
    {{\cal P} \left\{ z(t) = 0 \right\}}
    - 2\frac{{\cal P} \left\{ z(t) = 6 L \right\}}
    {{\cal P} \left\{ z(t) = 0 \right\}}
    + \ldots
\end{equation}
or
\begin{equation}
    {\cal P} \left\{ (\mbox{max} < L) \cap (\mbox{min} > -L) |z(t)=0\right\}
    = 1 + 2\sum_{n=1}^\infty (-1)^n \, \exp \left(-\frac{n^2L^2}{2t} \right)
\end{equation}
It has the same form as given by the method of images.

\subsection{The effect of destruction in the disk}

We now investigate the effect of destruction occurring in the disk, in the simple case where no boundary is 
present. As before, the steady-state 3-D solution will be obtained from the
one-dimensional time-dependent diffusion.
We denote $p$ the probability that a particle crossing the disk is destroyed.
In the case of a random walk $z=\sum_{i=1}^N z_i$ consisting of $N\equiv t/\tau$ 
elementary steps of length $\lambda$ and duration $\tau$, the probability distribution of
disk-crossing numbers $n$ is given by (Papoulis 2002)
\begin{equation}
    d{\cal P}_d (n | t) \equiv
    d{\cal P} (n | t, z(t) = 0 ) = \frac{2n \tau}{\kappa_2 t} \exp \left(
    - \frac{n^2 \tau}{\kappa_2 t} \right) \, dn\;.
\end{equation}
In this expression, $\kappa_2 \sim 1$ depends on its statistical properties 
(for instance, $\kappa_2 = 2$ for elementary steps
$z_i = \pm \lambda$ and $\kappa_2 \approx 1.43$ for $z_i$ uniformly
distributed in the interval $[-\lambda,\lambda]$).
The diffusion coefficient is defined as
\begin{equation}
    K \equiv \frac{\langle z^2 \rangle}{2t}
    = \frac{N\langle (z_i/\lambda)^2 \rangle}{2N} \frac{\lambda^2}{\tau}
    = \kappa_3 \frac{v^2 \tau}{2}\;,
\end{equation}
where $\kappa_3\equiv \langle (z_i/\lambda)^2 \rangle$ is the variance of
the elementary random step (in units of $\lambda$) and $v \equiv \lambda/\tau$. 
We thus finally have,
\begin{equation}
    d{\cal P}_d (n | t) = \frac{4Kn}{\kappa_2 \kappa_3 v^2
    t} \exp \left(
    - \frac{2Kn^2}{\kappa_2 \kappa_3 v^2 t}\right) \, dn
\end{equation}
We are now able to compute the probability distribution of disk
crossings for particles emitted from a distance $r$ in the disk as
\begin{equation}
    \frac{d{\cal P}_d (n | r)}{dn} = \int_0^\infty
    \frac{d{\cal P}_d (n | t)}{dn} \, {\cal P}_d (t | r) \,dt\;,
    \label{integrale_1}
\end{equation}
where the probability that a CR reaching distance $r$ in the disk
was emitted at time $t$ is
\begin{equation}
    {\cal P}_d (t | r) \propto \frac{1}{(Kt)^{3/2}}
    \exp \left( - \frac{r^2}{4K t}\right)\;.
\end{equation}
The above integral (\ref{integrale_1}) can be performed, yielding the
final result
\begin{equation}
    \frac{d{\cal P}_d (n | r)}{dn} = \frac{r_0^2 n}{r^2}
    \left( \displaystyle 1 + \frac{r_0^2 n^2}{r^2} \right)^{-3/2}\;,
    \label{final_nb_crossing}
\end{equation}
with
$r_0^2 \equiv 8 K^2/\kappa_2 \kappa_3 v^2
= 2 \lambda^2 \kappa_3/\kappa_2$.
We can also compute the integrated probability, that more that
$n_0$ crossings have
occurred, as
\begin{equation}
    {\cal P}_d (n>n_0 | r) =
    \left( \displaystyle 1 + \frac{r_0^2 n_0^2}{r^2}
    \right)^{-3/2} \;.
\end{equation}
A particle having crossed $n$ times the disk has the probability
$p_n = (1-p)^n \sim \exp(-np)$ of surviving, so that the survival
probability at distance $r$ is given by
\begin{eqnarray}
    {\cal P}_{\rm surv} (r) &=& \int_0^\infty \frac{d{\cal P} (n | r,
    z(t) = 0 )}{dn} \, e^{-np} \, dn \nonumber \\
    &=&
    \int_0^\infty \frac{x \, dx}{\left(1+x^2\right) ^{3/2}} \,
    e^{-xrp/r_0}\;. \nonumber
\end{eqnarray}
The density of Cosmic Rays in the disk is then given by
\begin{equation}
      N(r) = \frac{{\cal P}_{\rm surv} (r)}{4\pi K r}   
      = \frac{1}{4\pi K r} \int_0^\infty \frac{x \,
      dx}{\left(1+x^2\right) ^{3/2}} \,
      e^{- x r p/r_0}\;.
      \label{The_riri_equation}
\end{equation}
This form is actually equivalent to (\ref{sol_J0}), with $L \rightarrow \infty$,
\begin{equation}
    N(r,z=0) = \int_0^\infty \frac{k\, J_0(kr) \, dk}{2\pi(\Gamma + 2 K k)} 
\end{equation}
Indeed, rewriting $1/2\pi(\Gamma + 2 K k)$ as $\int_0^\infty dy\,
\exp(-2\pi(\Gamma + 2 K k)y)$ and reversing the integrations, 
(\ref{sol_J0_bis}) can be written,
\begin{equation}
    N(r,z=0) = \int_0^\infty dy \, e^{-2\pi\Gamma y} \int_0^\infty k\, J_0(kr) \, dk 
    e^{-4\pi K k y}
 \end{equation}
Using the identity (Gradshteyn \& Ryzhik 1980)
\begin{equation}
    \int_0^\infty e^{-\alpha y} J_0(\beta y) \, y \, dy =
    \frac{2 \alpha \beta}{(\alpha^2 + \beta^2)^{3/2}}
\end{equation}
we finally have, performing the change  of variables $4\pi Ky/r \equiv x$, 
\begin{equation}
    N(r,z=0) =  \int_0^\infty dy\, \frac{4\pi Kry}{(r^2 + (4\pi Ky)^2)^{3/2}} \,  e^{-2\pi\Gamma y}
    = \frac{1}{4 \pi Kr} \int_0^\infty dx\, \frac{x}{(1 + x^2)^{3/2}} \,  e^{-\Gamma rx/K\pi}
\end{equation}
This equation has  the same form as (\ref{The_riri_equation}), and relates the microscopic and macroscopic 
properties of diffusion, as $p/r_0$ must be equal to $\Gamma/\pi K$.
It should also be remarked that this integral is easier to compute than those involving $J_0$ functions, 
having a faster convergence as the integrand does not oscillate.
 
\section{Discussion}

The four methods presented in this paper do not all apply in every situation. 
The first two are valid for arbitrary $L$ and $\Gamma$ 
as well as with spontaneous decay of unstable species. 
The second one contains a richer physical information, as it provides 
the density as a function of time.
The third one is only valid when $\Gamma=0$, taking only into account 
the effect of absorption by the boundaries.
The fourth one is more general as we have also presented the case 
$\Gamma \neq 0$. However, it needs to be worked further out in order 
to consider simultaneously absorption at the boundaries and destruction 
in the central plane\footnote{For example, the probability 
distribution of disk crossing can be found in 
Taillet et al., {\tt astro-ph/0308141}, for a general case 
taking into account absorption, destruction and convective current}.
Comparison of the method of images and the use of random walk give 
some complementary insights to the consequences of absorption at a boundary. 
The paths that wander beyond the boundaries, and that we had to suppress 
by hand in Sec.~\ref{sec:random_walk} are actually those connecting 
the point at which the density is sought, to the negative images 
introduced in Sec.~\ref{sec:images}. The effect of these negative images is
to destroy the paths that would wander out of the diffusive volume.
It is important to note that this is different from the path integral
interpretation of the Schrodinger equation, in which case the paths
are weighted by a complex phase term, whereas in the classical diffusion
case we have discussed, the paths have only $\pm 1$ factors.


\begin{acknowledgments}
This work has benefited from the support of the PNC (Programme National de Cosmologie).
\end{acknowledgments}


\appendix

\section{Transformations of Eq. (21) for an easier numerical computation}
\label{demo}

\paragraph{Substraction} Using
\begin{equation}
      \int_0^\infty J_0(x) dx =1 
\end{equation}
$I[f]$ may be rewritten 
\begin{equation}
        I[f] =  1 - \int_0^\infty J_0(x) \, (1-f(x)) dx\;.
\end{equation}
The convergence is faster, as $1-f \rightarrow 0$ when $x
\rightarrow \infty$.

\paragraph{Integration by part} 
Using the identity $(xJ_1)' = xJ_0$ and integrating by parts,
one has
\begin{equation}
        I[f]
        = -\lim_{x \rightarrow 0^+} \left[ J_1(x) f(x) \right] 
	+\int_0^\infty J_1(x) \,
        \left( \frac{f(x)}{x} - f'(x)\right) dx\;.
\end{equation}
\onvire{\color{red} This expression is meaningful only if $x f(x)$ is
a bounded function near the origin $x=0$.}
Using the identity $J_0' = -J_1$ and integrating by parts again,
\begin{equation}
        I[f] =  -\lim_{x \rightarrow 0^+} \left[ J_1(x) f(x) \right] 
	+\left[J_0(x)\left(\frac{f(x)}{x}-f'(x)\right)
        \right]^\infty_0 
        +\int_0^\infty J_0(x) \,
        \left(f''(x) - \frac{f'(x)}{x} + \frac{f(x)}{x^2} \right)
        dx\;.
\end{equation}
\onvire{\color{red} The latter expression is meaningful only if $x^2 f(x)$ is
a bounded function near the origin $x=0$.}
These expressions provide several efficient alternatives 
to evaluate $I[f]$, provided the integrated terms are well defined, i.e. if
\begin{equation}
        f(x) = A + B x \ln x + {\cal O}(x)
\end{equation}


\paragraph{Comparison to a known function}

Part of the difficulty to evaluate numerically the Bessel expansions 
comes from the fact that the original functions are singular at the 
source position. As a result, the large $k$ modes continue to be important to reconstruct the solution.
We can take advantage of the fact that the singularity is known, as the
density is quite close to the free diffusion case $f_{\rm ref}(r)=1/4\pi Kr$ for small $r$.
The corresponding Bessel coefficients are given by
\begin{equation}
    \tilde{f}_{\rm ref}(k) = \int_0^\infty r \, \frac{1}{4 \pi K r} \, J_0(kr) dr
    = \frac{1}{4K\pi k}
\end{equation}
It is then judicious to write the density as
\begin{equation}
    N(r,0)= f_{\rm ref}(r) + \int_0^\infty k \, J_0(kr) \left\{ \tilde{f}(k)-\tilde{f}_{\rm ref}(k) \right\} \, dk
    = \frac{1}{4 \pi K r} + \int_0^\infty J_0(kr) \left\{ k \, \tilde{f}(k)-\frac{1}{4K\pi} \right\} \, dk
    \label{eq_ansatz}
\end{equation}
where the singularity is entirely contained in the first term:  
the Bessel expansion has been regularized. In the remaining integral to be computed, the convergence is much 
faster, as the large $k$ modes contribute very little.
Other choices of $f_{\rm ref}(r)$ may be preferred for particular values of $L$, $\Gamma$ and $r$.
This method yields a very good and rapid convergence for
sources located in the thin disk $z=0$.

\paragraph{Softening of the source term}

Finally, the source term may be spread out on a radius $a$, by
replacing the point source $\delta(\vec{r})$ by a disk source $q(r) = \theta(a-r)/(\pi a^2)$
for which an extra $2 J_1(k a)/ka$ term
appears in the Bessel transform.
With a judicious choice of the parameter $a$, the solution is very close 
to the original for $r\gg a$, but convergence is much faster due to the extra
$1/k$ factor.

\begin{figure}[!h]
    \centerline{\includegraphics[width=9cm]{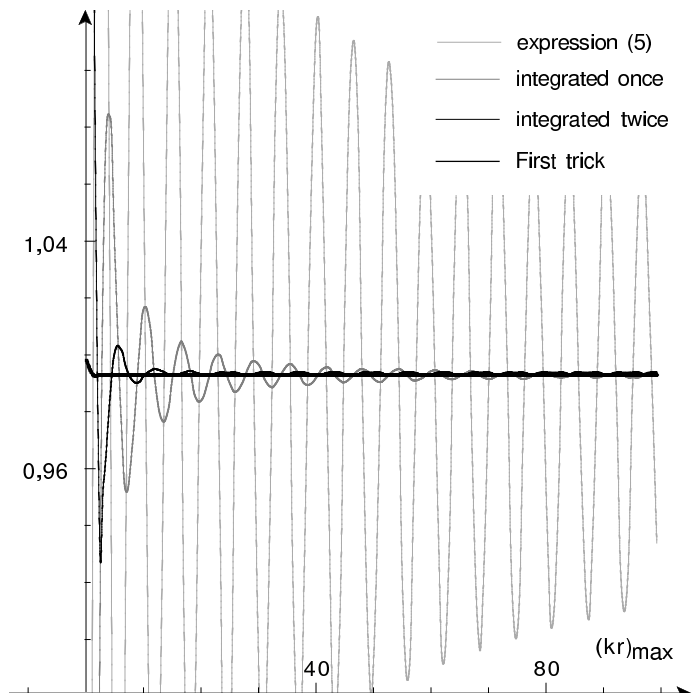}}
    \caption{Values of the correction to the free diffusion case, for $r=0.01\, L$ and $\Gamma=0$, computed 
    with the different methods described in the text, for different values of the upper boundary in the 
    integral.}
    \label{fig:test_convergence}
\end{figure}


\begin{thebibliography}{99}
\bibitem{taillet03} R. Taillet and D. Maurin, "{Spatial origin of Galactic cosmic rays in diffusion models. I. 
Standard sources in the Galactic disk}", {\it Astronomy and Astrophysics} {\bf 402}, 971-983 (2003).


\end{thebibliography}
\end{document}